\documentclass[12pt]{article}
\usepackage{amsthm}
\usepackage{eurosym}
\usepackage{amsfonts}
\usepackage{amssymb,amsmath}
\usepackage[dvips]{color}
\usepackage{amscd}
\usepackage{tikz}
\usepackage{mathtools}
\usepackage{graphicx}
\usepackage{caption}
\usepackage{subcaption}

\newcommand{\xdownarrow}[1]{ {\left\downarrow\vbox to #1{}\right.\kern-\nulldelimiterspace} }
\usetikzlibrary{decorations.pathreplacing}

\usetikzlibrary{patterns}

\def\and{\mathrm{and}}

\newcommand{\ee}{\end{equation}}
\newcommand{\bea}{\begin{eqnarray}}
\newcommand{\eea}{\end{eqnarray}}
\newcommand{\beas}{\begin{eqnarray*}}
\newcommand{\eeas}{\end{eqnarray*}}
\newcommand{\ba}{\begin{array}}
\newcommand{\ea}{\end{array}}

\newcommand{\nbox}{{\,\lower0.9pt\vbox{\hrule \hbox{\vrule height 0.2 cm \hskip 0.19 cm \vrule height 0.2 cm}\hrule}\,}}

\def\href#1#2{#2}
\theoremstyle{plain}

\begin{document}


\begin{titlepage}
\hfill
\vbox{
    \halign{#\hfil         \cr
           } 
      }  

\hbox to \hsize{{}\hss \vtop{ \hbox{}

}}

%

\vspace*{20mm}

\begin{center}

{\Large \textbf{Coherent state excitations and string-added coherent states in gauge-gravity correspondence\\ \vspace{0.4cm}}}

{\normalsize \vspace{10mm} }

{\normalsize {Hai Lin}}

{\normalsize \vspace{10mm} }

{\small \emph{\textit{Shing-Tung Yau Center of Southeast University, Southeast University,\\
Nanjing 210096, China
}} }

{\normalsize \vspace{0.2cm} }

{\small \emph{\textit{School of Mathematical Sciences, Southeast University, Nanjing 211189, China
}} }

{\normalsize \vspace{0.2cm} }

{\small \emph{\textit{Yau Mathematical Sciences Center, Tsinghua University,
Beijing 100084, China
\\
}} }
{\normalsize \vspace{0.4cm} }
\end{center}


\begin{abstract}

{\normalsize \vspace{0.3cm} }

  We analyze detailed properties of BPS coherent states and their connection to gravity. We interpret the group integral coherent state as a path integral over auxiliary variables coupled to the elementary letters of the theory. The eigenvalues of coherent state amplitudes can be viewed as collective coordinates of giant gravitons. Inspired by the above coherent states and by the integrability, we construct a new type of coherent states in the SL(2) sectors and their cousin PSU(1,1$|$2) sectors, analogous to the aforementioned coherent states. The large spin and small spin limits can be obtained by varying coherent state amplitudes. We add string words onto the BPS coherent states, and this gives rise to string-added coherent states. The insertions of string multi-words can be viewed as operator-insertions in this path integral. We describe BPS states and near BPS states building upon these coherent states in gauge-gravity correspondence. For example, string-added coherent states and their near BPS spectra are analyzed. This approach is particularly convenient for heavy excited states.

\end{abstract}

\end{titlepage}

\vskip 1cm

\section{Introduction}

\label{sec_introduction}

The gauge-gravity correspondence \cite%
{Maldacena:1997re,Gubser:1998bc,Witten:1998qj} is a nontrivial
correspondence between a quantum theory with gravity in the bulk and a
different quantum system on the boundary. The correspondence allows us to
perform calculations related to superstring theory and quantum gravity from
working on the quantum field theory side. On the other hand, the superstring
theory provides the UV completion of supergravity, and is a UV-complete
quantum gravity theory. Integrability \cite{Beisert:2010jr} has greatly
increased our understanding of the gauge-gravity correspondence. The
correspondence also reveals the nature of the emergent spacetime e.g. \cite%
{Rangamani:2016dms}-\cite{Balasubramanian:2005mg}. As we see, the bulk
emerges dynamically from the quantum mechanical description that lives in
fewer dimensions.

On the gravity side, giant graviton branes \cite{McGreevy:2000cw}-\cite%
{Berenstein:2004kk} are excited states. In the context of gauge-gravity
correspondence, emergent backreacted geometries correspond to highly excited
states in the quantum field theory side, such as the bubbling geometries
\cite{Lin:2004nb,Corley:2001zk,Berenstein:2004kk}. The states in the Hilbert
space of the quantum field theory are explicitly mapped to the gravity side.
Analysis in the field theory side shows that these different states live in
the same Hilbert space. The dual large operators and their representation
bases have been illuminated \cite{Corley:2001zk,Berenstein:2004kk}. These
heavy operators involve emergent backreacted geometries in the dual quantum
gravity system.

Here we focus on states which have interesting gravitational properties. One
interesting type of states are coherent states \cite%
{Berenstein:2022srd,Holguin:2022drf,Berenstein:2017abm}. Gravity dual of
coherent states has been analyzed. The coherent states in this paper are
related to the excited states of the gravitational spacetime. These set-ups
help to address the question how do curved spacetimes emerge from dual
quantum theory on the boundary. Coherent states also appear very widely in
many other contexts of physics, and here we concentrate on special types of
coherent states.

There are various representation bases for large operators. The large
operators includes those describe giant gravitons and backreacted emergent
geometries. The large operators also describe further excitations on these
heavy excited states. The correlation functions between light operators and
large operators can also be computed. The large operators can be expanded
systematically in terms of representation bases. In some sense, the larger the
operators are, there are more information that can be stored with the
operators. For more details along some of these ideas, see e.g. \cite%
{Ramgoolam:2008yr}-\cite{Kimura:2007wy},\cite{Berenstein:2005aa},\cite%
{Berenstein:2017abm},\cite{Berenstein:2022srd}. Various bases can be
transformed into each other. Operators of giant gravitons are also analyzed
by various important and insightful approaches \cite{Gaiotto:2021xce}-\cite%
{Lin:2012ey} and \cite{Berenstein:2022srd}. These approaches are closely
related to the scenarios needed for this paper. Various ideas have been put
forward, in order to make the computations with large operators more
efficient and convenient.

Coherent state operators in gauge-gravity duality have also been considered
in e.g. \cite%
{Simon:2018laf,Berenstein:2017abm,Lin:2017dnz,Berenstein:2022srd,Holguin:2022drf}
and their related references. These different bases have different labelings
due to their different symmetry properties of the multi-parameters of the
coherent states. In particular, \cite{Berenstein:2022srd} has manifest
permutation symmetry of the parameters. Mixed states and entangled states of
coherent state operators in the gauge-gravity correspondence have also been
considered \cite{Lin:2020qao}. Moreover, mixed states and entangled states
of Young tableau states were considered \cite{Lin:2021qso}. Many interesting
aspects of coherent states in relation to giant graviton states, were
explored in \cite{Berenstein:2022srd,Holguin:2022drf} and their related
references. The multi-parameter in the coherent states are collective
coordinates of giant gravitons.

The quantum gravitational system dual to these heavy excited state
operators, involves backreacted emergent geometries. Analysis on the field
theory side shows that these excited states live in the same Hilbert space
of the gravity side. Since they live in the same Hilbert space, we can
dynamically relate them using the Hamiltonian in the same Hilbert space. For
example one can superpose states and compute transition probabilities
between different states living in the gravity side, e.g. \cite{Brown:2006zk}%
-\cite{Mathur:2008kg},\cite{Berenstein:2017abm},\cite{Lin:2017dnz} and
related discussions.

When the spacetime is dynamical and quantum, the local quantum field theory
seems not adequate to describe these situations. On the other hand, string
theory is able to describe these situations very well. The set-ups discussed
in this paper can be put inside superstring theory, and the set-ups here are
UV-complete. Somehow we have gone beyond local quantum field theory.

In Sec. 2, we analyze detailed properties of coherent states and their
connection to gravity, among other things. We focus on above-mentioned heavy
states. Inspired by the above coherent states and by the integrability, in
Sec. 3, we construct a new type of coherent states in the $SL(2)$ sectors
and their cousins such as $PSU(1,1|2)$ sectors, among other things. The
large spin and small spin limits can be obtained by varying coherent state
amplitudes.

In Sec. 4. we analyze string-added coherent states and their near BPS
spectra, among other things. We can add some string words onto the BPS
coherent states, and this gives the string-added coherent states. We
identify and elaborate some classes of near BPS string-added coherent
states, whose anomalous dimension energy can be extracted conveniently. The
anormalous energy of the near BPS state is usually smaller than its bare
dimension energy. Hence these energy are controllable on both gauge theory
side and gravity side. One can expand around supersymmetric backgrounds to
compute their anormalous energy from both gauge theory and gravity side.
These states can be considered as excitations on the BPS background. This
method is very useful for near BPS states.

Finally in Sec. 5, we make conclusions and discuss some closely related
aspects. Note added: Upon the completion of this work, we received the
appearance of \cite{Holguin:2022drf} which have worked out $SO(N)$ and $%
Sp(N) $ cases; Many analysis in the current paper here also works for these
and other very interesting cases.

\section{BPS coherent states and collective variables}

\label{sec 2} \renewcommand{\theequation}{2.\arabic{equation}} %
\setcounter{equation}{0} \renewcommand{\thethm}{2.\arabic{thm}} %
\setcounter{thm}{0} \renewcommand{\theprop}{2.\arabic{prop}} %
\setcounter{prop}{0}

We consider a new class of BPS coherent states with manifest permutation
symmetries. This class of interesting coherent states in nonabelian gauge
theories have been constructed in \cite{Berenstein:2022srd}. Here, we first
analyze more detailed properties of the states. We use 4$D$ $\mathcal{N}=4$
gauge theory as an important concrete example. This theory is an example of
nonabelian gauge theories arising from multiple $D$-branes.

We begin by introducing some general set-ups. Operators $O~$acting on\ the
vacuum correspond to states $\left\vert \Psi \right\rangle ~$in the Hilbert
space, i.e. $\left\vert \Psi \right\rangle =O\left\vert 0\right\rangle $.
The operators can be built by the fundamental fields in the theory. For
example, in nonabelian gauge theories, we can have a complex matrix field $Z$%
, where we have some $U(N)$ gauge group. This $U(N)$ symmetry originates
from the symmetry on multiple $D$-branes. The field quanta are created and
annihilated by ladder operators. When acting on the operators, we have the
following correspondence
\begin{eqnarray}
Z &\leftrightarrow &a_{Z}^{\dagger },~~~~\partial _{Z}\leftrightarrow a_{Z}
\notag \\
Z_{j}^{i} &=&(Z)_{j}^{i}\leftrightarrow (a_{Z}^{\dagger })_{j}^{i}  \notag \\
(\partial _{Z})_{j}^{i} &=&\frac{\partial }{\partial (Z)_{i}^{j}}%
\leftrightarrow (a_{Z})_{j}^{i}
\end{eqnarray}%
The action $(a_{Z})_{j}^{i},$ $(\partial _{Z})_{j}^{i}$ is equivalent to
Wick contraction with $(Z)_{i}^{j}$. More detailedly, the right hand side of
the correspondence contains $\frac{1}{\sqrt{N}}$ norm factor for each
elementary field, and hence $\mathrm{Tr(}a_{Z}^{\dagger n})\left\vert
0\right\rangle $ corresponds to $\frac{1}{\sqrt{nN^{n}}}\mathrm{Tr}(Z^{n})$.
This is exact for any finite and fixed $N$. The convention here is that we
have the canonical commutation relations and normalizations for $%
a_{Z}^{\dagger },~a_{Z}.$

Similarly for other complex matrix fields $Y,X$ in the $U(N)$ gauge theory,
one has the correspondence
\begin{eqnarray}
Y &\leftrightarrow &a_{Y}^{\dagger },~~~~\partial _{Y}\leftrightarrow a_{Y}
\notag \\
X &\leftrightarrow &a_{X}^{\dagger },~~~~\partial _{X}\leftrightarrow a_{X}
\end{eqnarray}%
In the similar way, $\mathrm{Tr}(a_{Y}^{\dagger n})$ corresponds to $\mathrm{%
Tr}(Y^{n})$ and $\mathrm{Tr}(a_{X}^{\dagger n})$ corresponds to $\mathrm{Tr}%
(X^{n}).$

The new class of the BPS coherent states is
\begin{equation}
F[\Lambda ]=\frac{1}{Vol_{U}}\int dU\exp \left( \mathrm{Tr}(U\Lambda
U^{-1}a_{Z}^{\dagger })\right) \left\vert 0\right\rangle .  \label{coh_03}
\end{equation}%
Here $U\in U(N)$ is an unitary action on the $N$ $D$-branes. The action of
the unitary $U$, originates from the nonabelian gauge symmetry of the $N$
D-branes. The integral is over the group manifold, with the condition $\frac{%
1}{Vol_{U}}\int dU\cdot 1=1$. This defines a coherent state $\left\vert
F[\Lambda ]\right\rangle $. The conjugate bra state is $\langle F[\Lambda ]|$
and is defined by conjugating $a_{Z}^{\dagger }$ (respectively $\Lambda $)
to $a_{Z}$ (respectively $\bar{\Lambda}$).

We can also write the integral as $\int dg\exp (\mathrm{Tr}(\Lambda
(g^{-1}a_{Z}^{\dagger }g)))\left\vert 0\right\rangle $ and view $g$ as
auxiliary variables coupled to the $a_{Z}^{\dagger }$ fields. We then
integrate out these auxiliary variables. Here we interpret the integration
as a \textit{path integral} of the auxiliary variables. As a path integral,
it can be performed by a saddle point method.

The inner products of the states are
\begin{equation}
\langle F[\Lambda ^{\prime }]|F[\Lambda ]\rangle =\bar{F}[\bar{\Lambda}%
^{\prime }]\ast F[\Lambda ],
\end{equation}%
\begin{equation}
\bar{F}[\bar{\Lambda}^{\prime }]\ast F[\Lambda ]=\frac{1}{Vol}\int d\tilde{U}%
\exp (\mathrm{Tr}(\tilde{U}^{-1}\Lambda \tilde{U}\bar{\Lambda}^{\prime })).
\label{integral_07}
\end{equation}%
The normalizations of the states are $\mathcal{N}_{\Lambda }=\langle
F[\Lambda ]|F[\Lambda ]\rangle $, where
\begin{equation}
\mathcal{N}_{\Lambda }=\frac{1}{Vol}\int dU\exp (\mathrm{Tr}(U^{-1}\Lambda U%
\bar{\Lambda})).  \label{integral_08}
\end{equation}%
For more details, see \cite{Berenstein:2022srd}. Eqn. (\ref{integral_07}),(%
\ref{integral_08}) are Harish-Chandra-Itzykson-Zuber (HCIZ) integrals \cite%
{Harish,Itzykson:1979fi,Duistermaat:1982}, whose computation can be very
conveniently performed by localizations \cite{Duistermaat:1982} and by
saddle point methods.

We can define an unitary displacement operator
\begin{equation}
D(\Lambda )=\frac{1}{Vol_{U}}\int dU\exp \left( \mathrm{Tr}\left( U\Lambda
U^{-1}a_{Z}^{\dagger }-U\bar{\Lambda}U^{-1}a_{Z}\right) \right) .  \label{05}
\end{equation}%
Using Baker-Campbell-Hausdorff (BCH) formulas and commutation relations of
the ladder operators, we write%
\begin{equation}
F[\Lambda ]=\mathcal{N}_{\Lambda }^{\frac{1}{2}}D(\Lambda )\left\vert
0\right\rangle   \label{state 04}
\end{equation}%
and we show (\ref{coh_03}) and (\ref{state 04}) are equivalent. $D(\Lambda )$
is an unitary operation in $U(\mathcal{H})~$acting on the Hilbert space$~%
\mathcal{H}$ of states.\ In this writing, the term in the exponent is
manifestly anti-Hermitian, and hence the operation $D$ is an unitary
operation. The advantage is that it is manifestly unitary on the Hilbert
space. Using BCH formulas and commutation relations of the ladder operators,
we write $D$ also in the following way
\begin{equation}
D(\Lambda )=\frac{\mathcal{N}_{\Lambda }^{-\frac{1}{2}}}{Vol_{U}}\int dU\exp
\left( \mathrm{Tr}\left( U\Lambda U^{-1}a_{Z}^{\dagger }\right) \right) \exp
\left( -\mathrm{Tr}(U\bar{\Lambda}U^{-1}a_{Z})\right) .  \label{06}
\end{equation}%
Eqn. (\ref{05}) and (\ref{06}) are equivalent. The$~$state $F(\Lambda )$ is
an eigenstate of the annihilation operators and we have $\mathrm{Tr}%
(a_{Z}^{n})F(\Lambda )=\mathrm{Tr}(\Lambda ^{n})F(\Lambda )$ for gauge
invariant observables.

We can also define phase shift operator
\begin{equation}
R(\Theta _{Z})=\frac{1}{Vol}\int dU\exp (\mathrm{Tr}(iU\Theta
_{Z}U^{-1}a_{Z}^{\dagger }a_{Z})).  \label{03}
\end{equation}%
Here $\Theta _{Z}$ is a phase matrix, whose eigenvalues $\theta
_{z}^{i},i=1,...,N$,~are phases rotating the eigenvalues of$~\Lambda _{Z}$.
Here we let $\Lambda _{Z},\Theta _{Z}$ commute, i.e. $[\Lambda _{Z},\Theta
_{Z}]=0$. Hence
\begin{equation}
\frac{1}{Vol}\int dU\exp (\mathrm{Tr}(iU\Theta _{Z}U^{-1}a_{Z}^{\dagger
}a_{Z}))|F[\Lambda _{Z}]\rangle =\frac{1}{Vol}\int dU\exp (\mathrm{Tr}%
(U\Lambda _{Z}e^{i\Theta _{Z}}U^{-1}a_{Z}^{\dagger }))\left\vert
0\right\rangle
\end{equation}%
where we used BCH formulas. The eigenvalues are rotated by $\lambda
_{z}^{i}\rightarrow \lambda _{z}^{i}e^{i\theta _{z}^{i}}$, and
correspondingly, $\bar{\lambda}_{z}^{i}\rightarrow \bar{\lambda}%
_{z}^{i}e^{-i\theta _{z}^{i}}$.

The number operator is ${\hat{N}}_{Z}=a_{Z}^{\dagger }a_{Z}$.$~$We have ${%
\hat{J}}_{Z}=\mathrm{Tr}(a_{Z}^{\dagger }a_{Z})$ and ${J}_{Z}=\langle {\hat{J%
}}_{Z}\rangle $ measures the expectation value of the number of $Z$ fields
in the state.{\ The Hamiltonian operator is }${\hat{H}}_{0}=\mathrm{Tr}%
(a_{Z}^{\dagger }a_{Z})$ and $H_{0}=\langle {\hat{H}}_{0}\rangle $ counts
the excitation energy. Our Hamiltonian has subtracted out the Casimir energy
of the ground state. The Hamiltonian on the state space of the coherent
states is
\begin{equation}
E_{F[\Lambda ]}=\langle {\hat{H}}\rangle _{F[\Lambda ]}=\mathcal{N}%
_{F}^{-1}\langle F(\Lambda )|{\hat{H}}|F(\Lambda )\rangle =\mathcal{N}%
_{F}^{-1}~\bar{F}[\bar{\Lambda}]\ast {\hat{H}}\ast F[\Lambda ].
\end{equation}%
One can use a more sophisticated Lagrangian formalism \cite%
{Berenstein:2022srd} to obtain an effective action on coherent states
derived by \cite{Berenstein:2022srd}. These constructions and expressions
also works for the other matrix fields $Y,X$ in the $U(N)$ gauge theory by
using their ladder operators $a_{Y}^{\dagger },a_{X}^{\dagger }$ for (\ref%
{coh_03}).

The coherent states for coulomb branches also work similarly. Consider
coulomb branch gauge group $G_{1}\times G_{2}$ inside full gauge group $G$.
In that case, we need to embed $U(N_{1})\times U(N_{2})$ into $U(N)$, where $%
N_{1}+N_{2}=N,$ and the matrix $\Lambda _{Z}$ splits into blocks e.g. $%
\Lambda _{Z}^{(1)},\Lambda _{Z}^{(2)}$ corresponding to the two gauge
groups, and $\mathrm{rk}(\Lambda _{Z}^{(1)})+\mathrm{rk}(\Lambda _{Z}^{(2)})=%
\mathrm{rk}(\Lambda _{Z})$. Coulomb branch operators have been considered in
\cite{Diaz:2015tda} which is related to Gelfand-Tsetlin patterns \cite%
{Borodin Olshanski}. The integration $\int_{G}dU$ turns into $%
\int_{G_{1}\times G_{2}}\prod\limits_{i}dU_{i},$ where $U_{i}\in U(N_{i})$,$%
~i=1,2$. In this case, the permutation symmetry $S_{N}~$also reduces to $%
S_{N_{1}}\times S_{N_{2}}.$

Ref. \cite{Berenstein:2022srd} has also constructed a new class of eighth
BPS coherent states, with manifest permutation symmetries. It is constructed
by enlarging $U\Lambda _{Z}U^{-1}a_{Z}^{\dagger }$ in (\ref{coh_03}) to more
terms, adding additional parameters $\Lambda _{Y}$ and $\Lambda _{X}$. We
can take a limit case of the eighth BPS states constructed by \cite%
{Berenstein:2022srd}, by making $\Lambda _{X}\equiv 0$. For $\Lambda
_{(Z,Y)}\neq 0$, these are quarter BPS states. They are in the kernel of the
anormalous dimension dilatation operator, and this means $[\Lambda
_{Z},\Lambda _{Y}]=0$. This is a good advantage of using BPS coherent states.

These states are%
\begin{equation}
F[\Lambda _{Y},\Lambda _{Z}]=\frac{1}{Vol}\int dU\exp (\mathrm{Tr}(U\Lambda
_{Y}U^{-1}a_{Y}^{\dagger }+U\Lambda _{Z}U^{-1}a_{Z}^{\dagger }))\left\vert
0\right\rangle .  \label{coh_05}
\end{equation}%
Now first, in the following, we check that the states are quarter BPS at one
loop and two loop orders. The one-loop dilatation operator and effective
Hamiltonian is given by
\begin{equation}
\Delta _{2}=g_{YM}^{2}\mathrm{Tr}\left( [a_{Z}^{\dagger },a_{Y}^{\dagger
}][a_{Y},a_{Z}]\right) .
\end{equation}%
As pointed out by \cite{Berenstein:2022srd}, when we have the dilatation
operator act on $F$, we get a result that is equal to zero when the
parameters $\Lambda _{Y},\Lambda _{Z}$ are commuting matrices. The $%
a_{Y},a_{Z}$ appears on the rightmost. When acting on the above coherent
states, we have the identification $a_{Y}\leftrightarrow \Lambda _{Y}$,$%
~a_{Z}\leftrightarrow \Lambda _{Z}$, and $[a_{Y},a_{Z}]\leftrightarrow
\lbrack \Lambda _{Y},\Lambda _{Z}]=0$. Hence we see explicitly that the
action of the one-loop dilatation operator on (\ref{coh_05}) is zero.

The two-loop dilatation operator is
\begin{eqnarray}
\Delta _{4} &=&-\frac{g^{2}}{2}:\mathrm{Tr}\left( \left[ \left[
a_{Y}^{\dagger },a_{Z}^{\dagger }\right] ,a_{Z}\right] \left[ \left[
a_{Y},a_{Z}\right] ,a_{Z}^{\dagger }\right] \right) :  \notag \\
&&-\frac{g^{2}}{2}:\mathrm{Tr}\left( \left[ \left[ a_{Y}^{\dagger
},a_{Z}^{\dagger }\right] ,a_{Y}\right] \left[ \left[ a_{Y},a_{Z}\right]
,a_{Y}^{\dagger }\right] \right) :  \notag \\
&&-\frac{g^{2}}{2}:\mathrm{Tr}\left( \left[ \left[ a_{Y}^{\dagger
},a_{Z}^{\dagger }\right] ,T^{a}\right] \left[ \left[ a_{Y},a_{Z}\right]
,T^{a}\right] \right) :
\end{eqnarray}%
Here $g={\frac{g_{YM}^{2}}{8\pi ^{2}}}${\ with our convention.} The terms in
the dilatation operators are normal ordered. The normal ordering symbols
here indicate that the annihilation operators within the normal ordering
symbols do not act on fields inside the normal ordering. We computed the
action of the two-loop dilatation operator on the above coherent states, its
action on (\ref{coh_05}) is again zero, due to$~[\Lambda _{Z},\Lambda
_{Y}]=0 $. Then, by using nonrenormalization theorems, e.g. \cite%
{Lewis-Brown:2020nmg,Pasukonis:2010rv}, we are convinced that they are also
higher-loop BPS.

The state is a generating function of single-trace and multi-trace states of
the form $\mathrm{sTr}%
(Y^{m_{1}}Z^{n_{1}}Y^{m_{2}}Z^{n_{2}}...Y^{m_{l}}Z^{n_{l}}...)$. The
generated states are quarter BPS single-trace and multi-trace operators
built by $Z,Y$. They have been investigated in e.g. \cite%
{Lewis-Brown:2020nmg}. The two matrices $Z,Y$ correspond to $C^{2}$ of the
transverse dimensions of $N$ D-branes.

The unitary displacement operator is%
\begin{eqnarray}
D(\Lambda _{Y},\Lambda _{Z}) &=&\frac{1}{Vol_{U}}\int dU\exp (\mathrm{Tr}%
(U\Lambda _{Z}U^{-1}a_{Z}^{\dagger }+U\Lambda _{Y}U^{-1}a_{Y}^{\dagger })-
\notag \\
&&\mathrm{Tr}(U\bar{\Lambda}_{Z}U^{-1}a_{Z}+U\bar{\Lambda}_{Y}U^{-1}a_{Y}))
\label{08}
\end{eqnarray}%
and we write $F=\mathcal{N}_{F}^{\frac{1}{2}}D(\Lambda _{Y},\Lambda
_{Z})\left\vert 0\right\rangle $. It is equivalent to%
\begin{eqnarray}
D(\Lambda _{Y},\Lambda _{Z}) &=&\frac{\mathcal{N}_{F}^{-\frac{1}{2}}}{Vol_{U}%
}\int dU\exp (\mathrm{Tr}(U\Lambda U^{-1}a_{Z}^{\dagger }+U\Lambda
_{Y}U^{-1}a_{Y}^{\dagger }))  \notag \\
&&\exp (-\mathrm{Tr}(U\bar{\Lambda}U^{-1}a_{Z}+U\bar{\Lambda}%
_{Y}U^{-1}a_{Y})).
\end{eqnarray}

The phase shift operator is
\begin{equation}
R(\Theta _{Y})=\frac{1}{Vol}\int dU\exp (\mathrm{Tr}(iU\Theta
_{Y}U^{-1}a_{Y}^{\dagger }a_{Y})).  \label{04}
\end{equation}%
Here we let $\Lambda _{Y},\Theta _{Y}$ commute. The action of it rotates $%
\lambda _{y}^{i}\rightarrow \lambda _{y}^{i}e^{i\theta _{y}^{i}}.$

The Hamiltonian in the state space of coherent states is ${\hat{H}}_{0}=%
\mathrm{Tr}(a_{Y}^{\dagger }a_{Y}+a_{Z}^{\dagger }a_{Z})$. This counts the
BPS energy $H_{0}=\langle {\hat{H}}_{0}\rangle $. On the other hand, ${\hat{J%
}}_{Y}=\mathrm{Tr}(a_{Y}^{\dagger }a_{Y})$ and ${J}_{Y}=\langle {\hat{J}}%
_{Y}\rangle ~$measures the expectation value of the number of $Y$ fields in
the state.

Now we perform contracting the ladder operators and the inner product is%
\begin{equation}
\mathcal{N}_{F}=\bar{F}[\bar{\Lambda}_{Z},\bar{\Lambda}_{Y}]\ast F[\Lambda
_{Z},\Lambda _{Y}]=I(U,\Lambda _{\alpha },\bar{\Lambda}_{\alpha })
\end{equation}%
where $\alpha =Z,Y$. The integral is%
\begin{equation}
I(U,\Lambda _{\alpha },\bar{\Lambda}_{\alpha })=\frac{1}{Vol}\int dU\exp
\left( \Gamma _{\mathrm{HCIZ}}\left( U,\Lambda _{\alpha },\bar{\Lambda}%
_{\alpha }\right) \right) .  \label{integral_04}
\end{equation}%
We denote the exponent in (\ref{integral_04}) $\Gamma _{\mathrm{HCIZ}}$. The
exponent is%
\begin{equation}
\Gamma _{\mathrm{HCIZ}}\left( U,\Lambda _{\alpha },\bar{\Lambda}_{\alpha
}\right) =~\mathrm{Tr}(U\Lambda _{Y}U^{-1}\bar{\Lambda}_{Y}+U\Lambda
_{Z}U^{-1}\bar{\Lambda}_{Z}).  \label{exponent}
\end{equation}%
The path integral (\ref{integral_04}) can be computed by a saddle point
method. The saddle point equation is $d\Gamma _{\mathrm{HCIZ}}=0$. Now we
write $dK=UdU^{-1}$, and we have%
\begin{equation}
d\Gamma _{\mathrm{HCIZ}}=\ \mathrm{Tr}\left( dK[\bar{\Lambda}_{Y},U\Lambda
_{Y}U^{-1}]\right) +\mathrm{Tr}\left( dK[\bar{\Lambda}_{Z},U\Lambda
_{Z}U^{-1}]\right) =0.
\end{equation}%
The conditions for saddle points are%
\begin{equation}
\lbrack \bar{\Lambda}_{Y},U\Lambda _{Y}U^{-1}]+[\bar{\Lambda}_{Z},U\Lambda
_{Z}U^{-1}]=0.  \label{saddle}
\end{equation}%
When $U$ is a permutation matrix $P\in S_{N}$, these are saddle points. The
group integral can be viewed as a path integral of the auxiliary variables $%
U $. Then the integral (\ref{integral_04}) can be computed by localization
and saddle point method as described in \cite{Berenstein:2022srd}.

We now turn to the eighth BPS case in more details. The states are
\begin{equation}
F[\Lambda _{Z},\Lambda _{X},\Lambda _{Y}]=\frac{1}{Vol}\int dU\exp \left(
\mathrm{Tr}(U\Lambda _{X}U^{-1}a_{X}^{\dagger }+U\Lambda
_{Y}U^{-1}a_{Y}^{\dagger }+U\Lambda _{Z}U^{-1}a_{Z}^{\dagger })\right)
\left\vert 0\right\rangle .  \label{F_07}
\end{equation}%
It can also be written as
\begin{equation}
F[{\vec{\Lambda}}]=\frac{1}{Vol}\int dU\exp \left( \mathrm{Tr}(U{\vec{\Lambda%
}}U^{-1}{\vec{a}}^{\dagger })\right) \left\vert 0\right\rangle ,
\end{equation}%
where ${\vec{\Lambda}=(\Lambda _{X}},{\Lambda _{Y},\Lambda _{Z})}${, }${\vec{%
a}}^{\dagger }=(a_{X}^{\dagger },a_{Y}^{\dagger },a_{Z}^{\dagger })$. This
form of writing is also convenient for theories with global symmetries or
flavor symmetries. As pointed out in \cite{Berenstein:2022srd}, if the
parameters $\Lambda _{X},\Lambda _{Y},\Lambda _{Z}$ mutually commute, the
states are annihilated by the one-loop dilatation operator. The effective
Hamiltonian is given by
\begin{equation}
\Delta =g_{YM}^{2}\mathrm{Tr}([a_{X}^{\dagger },a_{Z}^{\dagger
}][a_{Z},a_{X}])+g_{YM}^{2}\mathrm{Tr}[a_{Y}^{\dagger },a_{Z}^{\dagger
}][a_{Z},a_{Y}])+g_{YM}^{2}\mathrm{Tr}([a_{X}^{\dagger },a_{Y}^{\dagger
}][a_{Y},a_{X}]).
\end{equation}%
When acting on the coherent states, we have the correspondence $%
a_{X}\leftrightarrow \Lambda _{X}$, $[a_{X},a_{Z}]\leftrightarrow \lbrack
\Lambda _{X},\Lambda _{Z}]$, $[a_{X},a_{Y}]\leftrightarrow \lbrack \Lambda
_{X},\Lambda _{Y}]$, etc. Hence all the three terms acting on the states are
zero, and the action of the dilatation is zero. Hence $\Lambda _{X},\Lambda
_{Y},\Lambda _{Z}$ are required to commute pairwise. By using
nonrenormalization theorems \cite{Lewis-Brown:2020nmg,Pasukonis:2010rv}, we
can infer that they are in the kernel of the anormalous dilatation operator.

The unitary displacement operator is similar to the above (\ref{08}) with
more terms in the exponent added. And we assume $\Lambda _{(X,Y,Z)}$ are
mutually commuting. By using BCH formulas, it is equivalent to the
following, with the normalization factor,
\begin{eqnarray}
&&D(\Lambda _{X},\Lambda _{Y},\Lambda _{Z})  \notag \\
&=&\frac{\mathcal{N}_{F}^{-\frac{1}{2}}}{Vol_{U}}\int dU\exp \left( \mathrm{%
Tr}(U\Lambda _{Z}U^{-1}a_{Z}^{\dagger }+U\Lambda _{X}U^{-1}a_{X}^{\dagger
}+U\Lambda _{Y}U^{-1}a_{Y}^{\dagger })\right)   \notag \\
&&\exp \left( -\mathrm{Tr}(U\bar{\Lambda}_{Z}U^{-1}a_{Z}+U\bar{\Lambda}%
_{X}U^{-1}a_{X}+U\bar{\Lambda}_{Y}U^{-1}a_{Y})\right) .
\end{eqnarray}%
We write $F=\mathcal{N}_{F}^{\frac{1}{2}}D(\Lambda _{X},\Lambda _{Y},\Lambda
_{Z})\left\vert 0\right\rangle .$ Similarly we define the phase shift
operator $R(\Theta _{X})~$similar to (\ref{03}) and (\ref{04}).

By performing usual manipulations contracting ladder operators, the overlap $%
\mathcal{N}_{F}~$was computed in \cite{Berenstein:2022srd}. The overlap is a
HCIZ integral, where the exponent is similar to (\ref{exponent}) and have
three terms, with $\Gamma _{\mathrm{HCIZ}}\left( U,\Lambda _{\alpha },\bar{%
\Lambda}_{\alpha }\right) =\mathrm{Tr}(U{\vec{\Lambda}}U^{-1}{\vec{\bar{%
\Lambda}}})$. The conditions for saddle points are in \cite%
{Berenstein:2022srd}. It is in a similar form as (\ref{saddle}) with three
terms. The integrals can be efficiently calculated by localization and
saddle point method.

The Hamiltonian in the space of coherent states is ${\hat{H}}_{0}=\mathrm{Tr}%
(a_{Z}^{\dagger }a_{Z}+a_{Y}^{\dagger }a_{Y}+a_{X}^{\dagger }a_{X})$, with
Casimir energy subtracted. The energy in the space of coherent states is%
\begin{equation}
{E}_{F}=\langle {\hat{H}}\rangle _{F}:=\mathcal{N}_{F}^{-1}\langle F[\Lambda
_{(X,Y,Z)}]|{\hat{H}}|F[\Lambda _{(X,Y,Z)}]\rangle =\mathcal{N}_{F}^{-1}~%
\bar{F}\ast H\ast F.
\end{equation}%
The angular momentum operator is ${\hat{J}}_{X}=\mathrm{Tr}(a_{X}^{\dagger
}a_{X})$ and ${J}_{X}=\langle {\hat{J}}_{X}\rangle $ measures the
expectation value of the number of $X$ fields in the state. We have that ${J}%
_{Z}=\mathrm{Tr}(\bar{\Lambda}_{Z}\Lambda _{Z})$,$~J_{Y}=\mathrm{Tr}(\bar{%
\Lambda}_{Y}\Lambda _{Y})$,$~J_{X}=\mathrm{Tr}(\bar{\Lambda}_{X}\Lambda
_{X}) $, and%
\begin{equation}
~H_{0}=J_{X}+J_{Y}+J_{Z}.
\end{equation}

The vevs are
\begin{eqnarray}
&&\langle \mathrm{Tr}(a_{Z}^{\dagger n}a_{Z}^{n})\rangle _{F}=\langle
\mathrm{Tr}(\bar{\Lambda}_{Z}\Lambda _{Z})^{n}\rangle _{F},~~~\langle
\mathrm{Tr}(a_{Y}^{\dagger n}a_{Y}^{n})\rangle _{F}=\langle \mathrm{Tr}(\bar{%
\Lambda}_{Y}\Lambda _{Y})^{n}\rangle _{F},~  \notag \\
&&\quad \quad \ \ \ \ \ \ \ \ \ \ \ \ \ \langle \mathrm{Tr}(a_{X}^{\dagger
n}a_{X}^{n})\rangle _{F}=\langle \mathrm{Tr}(\bar{\Lambda}_{X}\Lambda
_{X})^{n}\rangle _{F}.  \label{vev 05}
\end{eqnarray}%
Note that e.g. $(\bar{\Lambda}_{Z}\Lambda _{Z})^{n}$ is conveniently a
Hermitian matrix with real eigenvalues. We then look at the eigenvalue of
the parameters $\lambda _{i}^{x},\lambda _{i}^{y},\lambda _{i}^{z}$. In the
large $N$ limit, $\rho (\lambda )=\rho (\lambda ^{(x,y,z)})$ is an
eigenvalue density. The coherent state $|F[\Lambda _{(X,Y,Z)}]\rangle $ may
also be labeled as $|\rho (\lambda ^{(x,y,z)})\rangle $ in the large $N$
limit. We can also calculate the right hand sides of (\ref{vev 05}) for the
case when $n$ is not an integer, using eigenvalue formalism, e.g.
\begin{equation}
\langle \mathrm{Tr}(\bar{\Lambda}_{Z}\Lambda _{Z})^{n}\rangle _{F}=\int
d^{6}\lambda \rho (\lambda )(\bar{\lambda}^{z}\lambda ^{z})^{n}.
\end{equation}%
The methods developed in the approach of matrix eigenvalue effective models
are very useful in this context \cite{Berenstein:2005aa},\cite{Masuku:2009qf}%
-\cite{Berenstein:2008eg}. These vevs are dual to the multi-pole moments of
the gravitational geometries in the gravity side, and can be measured in the
gravity side via the methods of \cite{Skenderis:2007yb}-\cite{Liu:2007xj}.
The $n=1$ case of the similar quantities were analyzed in \cite{Chen:2007du}.

These states are heavy excited states in the gravity side. The particularly
interesting regimes are $J_{X},J_{Y},J_{Z}$ of order $N^{2}$ where there is
backreaction of the excitation onto the spacetime geometry. In these
regimes, single-trace and multi-traces are more difficult and less
convenient. The coherent states are especially convenient in their large
amplitude regimes. The gravity waves are collective excitations of these
large $N$ eigenvalue distributions. The order $N$ regimes describe multi
giant gravitons. The individual eigenvalue of the coherent state
multi-parameter is the collective coordinate of giant gravitons on the
gravity side \cite{Berenstein:2022srd}, see also related observations \cite%
{Berenstein:2017abm,Lin:2017dnz}.

This new type of coherent states have been constructed by \cite%
{Berenstein:2022srd,Holguin:2022drf}. The BPS coherent states are generating
functions of single-trace and multi-trace states. The multi-trace states are
multi closed string states. Hence the BPS coherent states are also
generating functions of multi closed string states. It is possible to
generalize to more matrices and more flavors, in other circumstances or in
other types of gauge theories. A special class of quarter BPS coherent
states were also constructed in \cite{Lin:2017vfn}, which has not made use
of the manifest permutation symmetries.

\section{The $SL(2)$ sectors, their cousins, and meromorphic versions}

\label{sec 3} \renewcommand{\theequation}{3.\arabic{equation}} %
\setcounter{equation}{0} \renewcommand{\thethm}{3.\arabic{thm}} %
\setcounter{thm}{0} \renewcommand{\theprop}{3.\arabic{prop}} %
\setcounter{prop}{0}

We construct a new type of coherent states in the $SL(2)$ sectors and their
cousin $PSU(1,1|2$) sectors. We also construct a meromorphic version of
coherent states, which can be transformed back to the full coherent states
when $\det \Lambda _{Z}\neq 0$. We give transformation relations between
different types of coherent states, among other things.

Inspired by the coherent states in \cite{Berenstein:2022srd}, whose aspects
are analyzed in more details in Sec. 2, we construct a coherent state as
generating function of%
\begin{equation}
\mathrm{Tr}(D_{+}^{n_{1}}ZD_{+}^{n_{2}}Z...D_{+}^{n_{\alpha }}Z...),
\label{states_15}
\end{equation}%
$n_{\alpha }=0,1,2,...$ and so on. Here $D_{+}$ is a light-cone covariant
derivative operator. We write $D_{+}=n^{\mu }D_{\mu }$ where $n^{\mu }$ is a
light-like vector. $D_{+}$ is in the (1,0) representation of the local
Lorentz group. When $D_{+}$ are dilute, they can be viewed as impurities on
top of $Z$. Similar conceptions are also raised in \cite%
{deMelloKoch:2011vn,Berenstein:2020jen}; see also \cite{Kim:2018gwx}. For
more details on $SL(2)$ sectors and their relations to integrability and
QCD, see for example \cite{Beisert:2010jr,Tseytlin,Belitsky:2003ys}.

We can construct the ladder operators corresponding to adding derivatives $n$
times on top of $Z$,%
\begin{equation}
c_{D}^{\dagger n}a_{Z}^{\dagger }\leftrightarrow
D_{+}^{n}Z,~~~c_{D}^{\dagger }a_{Z}^{\dagger }\leftrightarrow
D_{+}Z,~~~a_{Z}^{\dagger }\leftrightarrow Z.
\end{equation}%
We construct a type of coherent states as the generating function of (\ref%
{states_15}),%
\begin{equation}
K[R_{Z},\Lambda _{Z}]=\frac{1}{Vol}\int dU\exp (\mathrm{Tr}%
(UR_{Z}U^{-1}c_{D}^{\dagger }a_{Z}^{\dagger }a_{Z}))\exp (\mathrm{Tr}%
(U\Lambda _{Z}U^{-1}a_{Z}^{\dagger }))\left\vert 0\right\rangle .
\label{states_17}
\end{equation}%
Here $R_{Z},\Lambda _{Z}$ commute. We make sure that before acting $%
c_{D}^{\dagger }$, there is already a background of $Z$-fields due to the
right-most operator in (\ref{states_17}). The angular momentum operator and
spin operator are ${\hat{J}}_{Z}=\mathrm{Tr}(a_{Z}^{\dagger }a_{Z})$ and ${%
\hat{S}}=\mathrm{Tr}(c_{D}^{\dagger }c_{D})$. Here $S=\langle {\hat{S}}%
\rangle $ measures the expectation value of the number of $D_{+}$. On this
state (\ref{states_17}), the expectation values are $J_{Z}=\langle {\hat{J}}%
_{Z}\rangle =\mathrm{Tr}({\bar{\Lambda}}_{Z}\Lambda _{Z})$ and $S=\langle {%
\hat{S}}\rangle =\mathrm{Tr}({R}_{Z}{\bar{R}}_{Z}{\bar{\Lambda}}%
_{Z}^{2}\Lambda _{Z}^{2})$.

An alternative formulation is the state%
\begin{equation}
G[T_{DZ},\Lambda _{Z}]=\frac{1}{Vol}\int dU\exp (\mathrm{Tr}%
(UT_{DZ}U^{-1}c_{D}^{\dagger }a_{Z})\exp (\mathrm{Tr}(U\Lambda
_{Z}U^{-1}a_{Z}^{\dagger }))\left\vert 0\right\rangle .  \label{states_G_03}
\end{equation}%
Here $T_{DZ},\Lambda _{Z}$ commute. We have that $J_{Z}=\mathrm{Tr}({\bar{%
\Lambda}}_{Z}\Lambda _{Z})$ and $S=\mathrm{Tr}({T}_{DZ}{\bar{T}}_{DZ}{\bar{%
\Lambda}}_{Z}\Lambda _{Z}).$

These are heavy states in the gravity side. If the $D_{+}$ is very dilute,
it is the small spin limit. This is when $\alpha =N^{-1}\mathrm{Tr}({T}_{DZ}{%
\bar{T}}_{DZ})$ is much smaller than one. On the other hand, if the $Z$ is
dilute and $D_{+}$ is dense, it is the large spin limit. This is when $%
\alpha $ is much larger than one. The large spin limit has been discussed
also in \cite{deMelloKoch:2011vn}.

States of this type (\ref{states_17}) can also be constructed for the
operators analyzed in sections 2. We construct the following state
\begin{equation}
G[T_{Y},\Lambda _{Z}]=\frac{1}{Vol}\int dU\exp (\mathrm{Tr}%
(UT_{Y}U^{-1}a_{Y}^{\dagger }a_{Z}))\exp (\mathrm{Tr}(U\Lambda
_{Z}U^{-1}a_{Z}^{\dagger }))\left\vert 0\right\rangle .  \label{states_G_05}
\end{equation}%
Here $T_{Y},\Lambda _{Z}$ commute. The right-most operator in (\ref%
{states_G_05}) is an eigenstate of $a_{Z}$. When acting on the coherent
state background, we replace the annihilation operator by the vev. Here we
replace the $a_{Z}~$by vev $\Lambda _{Z}$, when the left operator in (\ref%
{states_G_05}) acts on the right-most operator. The action $a_{Y}^{\dagger
}a_{Z}$ annihilates a $Z$ field and substitute it with a $Y$ field, e.g. $%
Z^{2}\rightarrow YZ$. Hence we have $J_{Z}=\mathrm{Tr}({\bar{\Lambda}}%
_{Z}\Lambda _{Z})$ and $J_{Y}=\mathrm{Tr}({T}_{Y}{\bar{T}}_{Y}{\bar{\Lambda}}%
_{Z}\Lambda _{Z}).$

The state (\ref{states_G_05}) is a meromorphic version of (\ref{coh_05}). $%
G[T_{Y},\Lambda _{Z}]$ is the excitation on top of $F[\Lambda _{Z}]$. It is
a good approximation to $F[\Lambda _{Y},\Lambda _{Z}]$, in the regime $\det
\Lambda _{Z}\neq 0.$ We see that
\begin{equation}
G[T_{Y},\Lambda _{Z}]|_{T_{Y}=\Lambda _{Z}^{-1}\Lambda _{Y}}\sim F[\Lambda
_{Y},\Lambda _{Z}],
\end{equation}%
when $\det \Lambda _{Z}\neq 0$ and identifying $T_{Y}=\Lambda
_{Z}^{-1}\Lambda _{Y}$. When $\det \Lambda _{Z}=0,$ $G[T_{Y},\Lambda _{Z}]$
can not be transformed back to $F[\Lambda _{Y},\Lambda _{Z}],$ since there
is pole in $T_{Y}$. Here $T_{Y}$ is a meromorphic function of $\Lambda _{Z}$
and not a holomorphic function, hence there is pole in $T_{Y}.$ This is when
$\det \Lambda _{Z}=0$. $F[\Lambda _{Y},\Lambda _{Z}]$ is a smooth resolution
of $G[T_{Y},\Lambda _{Z}]$, extending it to cover the pole locus $\det
\Lambda _{Z}=0.$

The meromorphic version provides new understandings of the coherent states
and new understanding to the important question why we have the background
fields $Z$. These background fields, e.g. $Z$ fields, serve as new effective
vacuum for new excitations.

Similarly, for including $X$, we have the state%
\begin{eqnarray}
&&G[T_{X},T_{Y},\Lambda _{Z}]=  \notag \\
&&\frac{1}{Vol}\int dU\exp (\mathrm{Tr}(UT_{X}U^{-1}a_{X}^{\dagger
}a_{Z}+UT_{Y}U^{-1}a_{Y}^{\dagger }a_{Z})\exp (\mathrm{Tr}(U\Lambda
_{Z}U^{-1}a_{Z}^{\dagger }))\left\vert 0\right\rangle .  \notag \\
&&  \label{states_G_07}
\end{eqnarray}%
Here $T_{X},T_{Y},\Lambda _{Z}$ mutually commute. We have that $J_{Z}=%
\mathrm{Tr}({\bar{\Lambda}}_{Z}\Lambda _{Z}),J_{Y}=\mathrm{Tr}({T}_{Y}{\bar{T%
}}_{Y}{\bar{\Lambda}}_{Z}\Lambda _{Z})$, and $J_{X}=\mathrm{Tr}({T}_{X}{\bar{%
T}}_{X}{\bar{\Lambda}}_{Z}\Lambda _{Z}).$

In the regime $\det \Lambda _{Z}\neq 0,$%
\begin{equation}
G[T_{X},T_{Y},\Lambda _{Z}]\sim F[\Lambda _{X},\Lambda _{Y},\Lambda _{Z}],
\end{equation}%
with $T_{X}=\Lambda _{Z}^{-1}\Lambda _{X},T_{Y}=\Lambda _{Z}^{-1}\Lambda
_{Y} $. Hence state (\ref{states_G_07}) is a meromorphic version of (\ref%
{F_07}). $G[T_{Y},\Lambda _{Z}]$, $G[T_{X},T_{Y},\Lambda _{Z}]$ can be
viewed as meromorphic versions of quarter and eighth BPS coherent states.

Here, it is very similar to having $U(2)$ and $U(3)$ flavor symmetry, or
more generally $U(N_{F})$ symmetry if there are $N_{F}~$sets of fields with
global symmetry. Similar expansions also works for quiver gauge theory with
global symmetry, as well as their Coulomb branches. On the other hands,
there are theories with these flavor symmetries.

Turning to the $SL(2)$ sector, this sector can be enlarged to $PSU(1,1|2$)
sector. The $PSU(1,1|2)$ sector has been discussed from the point of view of
integrability, e.g. \cite{Beisert:2007sk}. Now we consider bosonic part of $%
PSU(1,1|2)$ sector, which is composed of $Z,Y$, $D_{+}$ and two fermions in
the ($\frac{1}{2}$, -$\frac{1}{2}$) and ($\frac{1}{2}$, $\frac{1}{2}$) of
the local Lorentz group. In the $PSU(1,1|2)$ sector, the bosonic part of the
state is%
\begin{eqnarray}
G[T_{DZ},T_{DY},\Lambda _{Z},\Lambda _{Y}] &=&\frac{1}{Vol}\int dU\exp (%
\mathrm{Tr}(UT_{DZ}U^{-1}c_{D}^{\dagger }a_{Z})+\mathrm{Tr}%
(UT_{DY}U^{-1}c_{D}^{\dagger }a_{Y}))  \notag \\
&&\exp (\mathrm{Tr}(U\Lambda _{Z}U^{-1}a_{Z}^{\dagger }+U\Lambda
_{Y}U^{-1}a_{Y}^{\dagger })\left\vert 0\right\rangle .
\end{eqnarray}%
Here $T_{DZ},T_{DY},\Lambda _{Z},\Lambda _{Y}$ mutually commute. The angular
momentum and spin of the state is $J_{Z}=\mathrm{Tr}({\bar{\Lambda}}%
_{Z}\Lambda _{Z}),J_{Y}=\mathrm{Tr}({\bar{\Lambda}}_{Y}\Lambda _{Y})$ and $S=%
\mathrm{Tr}(|T_{DZ}\Lambda _{Z}+T_{DY}\Lambda _{Y}|^{2}).$

An alternative formulation in the bosonic part of the $PSU(1,1|2)$ sector, is%
\begin{eqnarray}
K[R_{Z},R_{Y},\Lambda _{Z},\Lambda _{Y}] &=&\frac{1}{Vol}\int dU\exp (%
\mathrm{Tr}(UR_{Z}U^{-1}c_{D}^{\dagger }a_{Z}^{\dagger }a_{Z})+\mathrm{Tr}%
(UR_{Y}U^{-1}c_{D}^{\dagger }a_{Y}^{\dagger }a_{Y}))  \notag \\
&&\exp (\mathrm{Tr}(U\Lambda _{Z}U^{-1}a_{Z}^{\dagger }+U\Lambda
_{Y}U^{-1}a_{Y}^{\dagger })\left\vert 0\right\rangle .
\end{eqnarray}%
Here $R_{Z},\Lambda _{Z},R_{Y},\Lambda _{Y}$ mutually commute.$~$We see that
$J_{Z}=\mathrm{Tr}({\bar{\Lambda}}_{Z}\Lambda _{Z}),J_{Y}=\mathrm{Tr}({\bar{%
\Lambda}}_{Y}\Lambda _{Y})$ and $S=\mathrm{Tr}(|R_{Z}{\bar{\Lambda}}%
_{Z}\Lambda _{Z}+R_{Y}{\bar{\Lambda}}_{Y}\Lambda _{Y}|^{2})$.

Now we turn to the quiver case. For quiver gauge theories and orbifold
daughters of $\mathcal{N}=4$ theory, the coherent state construction is
similar. Many orbifold daughters have integrability. Many aspects of them
have been discussed and worked out in e.g. \cite{Berenstein:2006yy}-\cite%
{Pasukonis:2013ts} and their related references. For example, we can make a
projection of $U(MN)$ theory to get the $U(N)^{M}$ theory. The construction
of coherent states for quiver gauge theories have been worked out in \cite%
{Berenstein:2022srd}. For example, we have a $U(N_{1})\times U(N_{2})$ gauge
group. We consider a pair of bifundamental fields $a_{12}^{\dagger
},a_{21}^{\dagger }$ in the $(\bar{N}_{1},N_{2})$ and the $(N_{2},\bar{N}%
_{1})$ representations. We make the same type of coherent states with their
parameters $\Lambda _{21},\Lambda _{12}$, whose roles are similar to $%
\Lambda _{Z},\Lambda _{Y}$. The state is%
\begin{equation}
F[\Lambda _{21},\Lambda _{12}]=\frac{1}{Vol}\int \prod_{i=1}^{2}dU_{i}\exp
\left( \mathrm{Tr}(\Lambda _{21}U_{1}a_{12}^{\dagger }U_{2}^{-1}+\Lambda
_{12}U_{2}a_{21}^{\dagger }U_{1}^{-1}\right) \left\vert 0\right\rangle .
\label{states_18}
\end{equation}%
Note that the conjugate of $a_{21}^{\dagger }$ is $a_{12}$, and the
conjugate of $a_{12}^{\dagger }$ is $a_{21}$, with our convention. Now we
construct a new state, which is the meromorphic version of (\ref{states_18}),%
\begin{equation}
G[T_{21},\Lambda _{12}]=\frac{1}{Vol}\int \prod_{i=1}^{2}dU_{i}\exp (\mathrm{%
Tr}(U_{2}^{-1}T_{21}U_{1}a_{12}^{\dagger }a_{12}))\exp (\mathrm{Tr}%
(U_{1}^{-1}\Lambda _{12}U_{2}a_{21}^{\dagger }))\left\vert 0\right\rangle .
\label{G_19}
\end{equation}%
~The right-most operator in (\ref{G_19}) is an eigenstate of $a_{12}$. We
see that
\begin{equation}
G[T_{21},\Lambda _{12}]\sim F[\Lambda _{21},\Lambda _{12}],
\end{equation}%
with $\Lambda _{12}T_{21}=\Lambda _{21}$, $T_{21}=\Lambda _{12}^{-1}\Lambda
_{21}$. $G[T_{21},\Lambda _{12}]$ is a meromorphic version of $F[\Lambda
_{21},\Lambda _{12}]$ and can be transformed to it when $N_{1}=N_{2}$ and $%
\det \Lambda _{12}\neq 0$. The inner products of (\ref{states_18}) have been
computed in \cite{Berenstein:2022srd}, where localization techniques have
been implemented.

\section{String-added coherent states}

\label{sec 4} \renewcommand{\theequation}{4.\arabic{equation}} %
\setcounter{equation}{0} \renewcommand{\thethm}{4.\arabic{thm}} %
\setcounter{thm}{0} \renewcommand{\theprop}{4.\arabic{prop}} %
\setcounter{prop}{0}

The BPS coherent state can serve as the supersymmetric background for other
excitations on top of it. We can consider further excitations on the BPS
coherent state background, by adding strings on top of them. One important
idea is splitting and identification of fields for the background and fields
for the impurities. We begin with some general settings.

\subsection{General settings with multi-words}

The background BPS coherent state is (\ref{F_07}). On this background, other
fields are excited. This is to add words or multi-words onto the coherent
state operator. These other excited fields may be viewed as impurities, if
their numbers are much fewer than the number of background fields. We
generalize the ideas of \cite{Berenstein:2022srd} from single word to
multi-words, and from single pair to multi-pairs.

The BPS coherent states are parametrized by $\vec{\Lambda}=(\Lambda
_{X},\Lambda _{Y},\Lambda _{Z})$. Consider $\left\vert v_{x}\right\rangle
_{i}$, $\left\vert v_{y}\right\rangle _{i}$, $\left\vert v_{z}\right\rangle
_{i}$ are eigenvectors, i.e.%
\begin{equation}
\Lambda _{X}\left\vert v_{x}\right\rangle _{i}=\lambda _{x}^{i}\left\vert
v_{x}\right\rangle _{i},\ ~\Lambda _{Y}\left\vert v_{y}\right\rangle
_{i}=\lambda _{y}^{i}\left\vert v_{y}\right\rangle _{i},~~\Lambda
_{Z}\left\vert v_{z}\right\rangle _{i}=\lambda _{z}^{i}\left\vert
v_{z}\right\rangle _{i}.
\end{equation}%
When we multiply as $U\Lambda _{X}U^{-1}$, $U\Lambda _{Y}U^{-1},$ $U\Lambda
_{Z}U^{-1}$,$~$the vectors are rotated to $U\left\vert v_{x}\right\rangle
_{i}$, $U\left\vert v_{y}\right\rangle _{i},$ $U\left\vert
v_{z}\right\rangle _{i}$. The bra states are rotated to $\langle
v_{x}|_{i}U^{-1}$, etc. The vectors are simultaneously rotated. There is
also a global symmetry $U(3)$ acting on the set of 3 vectors.

The analysis of \cite{Berenstein:2022srd} as well as the above analysis in
Sec. 2 imply that $\Lambda _{(X,Y,Z)}$ are mutually commuting for BPS. Now
we assume they mutually commute. Hence they can share the same
eigenvectors.\ We consider there are two large eigenvalues $\lambda
_{(X,Y,Z)}^{i}$ and $\lambda _{(X,Y,Z)}^{j}$. They share the same
eigenvectors,$~$%
\begin{eqnarray}
\Lambda _{X}\left\vert \vec{v}\right\rangle _{\alpha } &=&\lambda
_{x}^{\alpha }\left\vert \vec{v}\right\rangle _{\alpha },~\ \Lambda
_{Y}\left\vert \vec{v}\right\rangle _{\alpha }=\lambda _{y}^{\alpha
}\left\vert \vec{v}\right\rangle _{\alpha },~~~\Lambda _{Z}\left\vert \vec{v}%
\right\rangle _{\alpha }=\lambda _{z}^{\alpha }\left\vert \vec{v}%
\right\rangle _{\alpha },  \notag \\
\mathrm{for}\text{ }\alpha &=&i,j.
\end{eqnarray}%
Here only for two eigenvalues $i,j$ having this property are needed. Hence $%
\left\vert v_{x}\right\rangle _{\alpha }=\left\vert v_{y}\right\rangle
_{\alpha }=\left\vert v_{z}\right\rangle _{\alpha }=\left\vert \vec{v}%
\right\rangle _{\alpha },$ $\mathrm{for}$ $\alpha =i,j.$ This means that in
the two dimensional subspace involving $i,j$, $\Lambda _{(X,Y,Z)}$ share the
same eigenvector $\left\vert \vec{v}\right\rangle _{i,j}$. \

The added words on top is
\begin{equation}
~\ \langle \vec{v}|_{i}U^{-1}WU\left\vert \vec{v}\right\rangle _{j}~,~
\end{equation}%
and the state is%
\begin{equation}
\frac{1}{Vol}\int dU\exp \left( \mathrm{Tr}(U\Lambda
_{X}U^{-1}a_{X}^{\dagger }+U\Lambda _{Y}U^{-1}a_{Y}^{\dagger }+U\Lambda
_{Z}U^{-1}a_{Z}^{\dagger })\right) \langle \vec{v}|_{i}U^{-1}WU\left\vert
\vec{v}\right\rangle _{j}\left\vert 0\right\rangle .  \label{words_02}
\end{equation}%
More generally, we add multi-words of this kind,%
\begin{eqnarray}
&&\frac{1}{Vol}\int dU\exp \left( \mathrm{Tr}(U\Lambda
_{X}U^{-1}a_{X}^{\dagger }+U\Lambda _{Y}U^{-1}a_{Y}^{\dagger }+U\Lambda
_{Z}U^{-1}a_{Z}^{\dagger })\right) \cdot  \notag \\
&&\langle \vec{v}|_{i_{1}}U^{-1}W_{1}U\left\vert \vec{v}\right\rangle
_{j_{1}}\langle \vec{v}|_{i_{2}}U^{-1}W_{2}U\left\vert \vec{v}\right\rangle
_{j_{2}}...\langle \vec{v}|_{i_{l}}U^{-1}W_{l}U\left\vert \vec{v}%
\right\rangle _{j_{l}}\left\vert 0\right\rangle .  \label{words_multiple_02}
\end{eqnarray}%
The indices $(i_{l},j_{l})\in I_{s}$ form a set $I_{s}$. $W$ is a word, in
other words, a string. For example, $W$ is schematically$\ D_{+}^{\dagger
n_{1}}a_{Z}^{\dagger n_{2}}a_{Y}^{\dagger n_{3}}a_{X}^{\dagger n_{4}}$, with
$n_{\alpha }=0,1,2,...$, and so on. $W$ can be labelled by BMN operators or
spin chain like operators, and we can also include $a_{{\Bar{Z}}}^{\dagger }$
etc.

The integral (\ref{words_multiple_02}) can be viewed as a path integral of
the auxiliary variable $U$ which is coupled to $a_{Z}^{\dagger
},a_{Y}^{\dagger },a_{X}^{\dagger }$ and $W_{\alpha }$, in accord with the
alternative interpretation in Sec 2. The insertions of $W_{\alpha }$ can be
viewed as operator-insertion in this path integral (\ref{words_multiple_02}%
). We have generalized the word insertions in (\ref{words_02}) to multi-word
insertions, and interpreted them as operator insertions in the path integral
of auxiliary variables.

We extract the anormalous energy from the combined state, e.g. $F\cdot W$.
The background energy are energy of BPS states. We extract the energy of
non-BPS part of the excitation, corresponding to the anormalous dimension.
The anomalous dimensions give rise to non-BPS energy on top of BPS energy.
We works on the anormalous energy due to the quartic interaction vertices.
The quartic interaction vertex connects one impurity field on the word part
and one background field on the coherent state part, and they then Wick
contract to their conjugates respectively. We can also use the dilatation
operator.

The simplest cases are when the coherent state part $F$ and the word part $W$
are both BPS. The coherent state background $F$ is BPS, hence there is no
non-BPS energy from the interaction inside the coherent state part. Then the
anomalous dimension energy is solely due to the interaction between the
coherent state part and the word part. One of the simple cases is that $W$
is an eighth BPS string. The entire state can be maintained as a near BPS
state. Sec. 4.2 are special cases of these ideas.

\subsection{Near BPS string-added coherent states}

We first consider near BPS states of the form%
\begin{eqnarray}
&&\frac{1}{Vol}\int dU\exp \left( \mathrm{Tr}(U\Lambda
_{Z}U^{-1}a_{Z}^{\dagger })\right) \cdot  \notag \\
&&\langle \vec{v}|_{i_{1}}U^{-1}W_{1}U\left\vert \vec{v}\right\rangle
_{j_{1}}\langle \vec{v}|_{i_{2}}U^{-1}W_{2}U\left\vert \vec{v}\right\rangle
_{j_{2}}...\langle \vec{v}|_{i_{l}}U^{-1}W_{l}U\left\vert \vec{v}%
\right\rangle _{j_{l}}\left\vert 0\right\rangle .  \label{state 08}
\end{eqnarray}%
The entire state can be maintained as a near BPS state. For example we add
impurities corresponding to $X$, and the word $W~$is $a_{X}^{\dagger
}a_{Z}^{\dagger m}$, $m=0,1,...~$and so on. In this case, $W$ is BPS itself,
and it can be viewed as a symmetrized state in the sector of $Z,X$. The
anormalous energy is the energy corresponding to the anormalous dimension.
Because $W$ and $F$ both has no anormalous dimension under the dilatation
operator, the anormalous dimension energy is due to the interaction between
the coherent state $F$ part and the word $W$ part.

We then consider near BPS states of the form%
\begin{eqnarray}
&&\frac{1}{Vol}\int dU\exp \left( \mathrm{Tr}(U\Lambda
_{Z}U^{-1}a_{Z}^{\dagger }+U\Lambda _{Y}U^{-1}a_{Y}^{\dagger })\right) \cdot
\notag \\
&&\langle \vec{v}|_{i_{1}}U^{-1}W_{1}U\left\vert \vec{v}\right\rangle
_{j_{1}}\langle \vec{v}|_{i_{2}}U^{-1}W_{2}U\left\vert \vec{v}\right\rangle
_{j_{2}}...\langle \vec{v}|_{i_{l}}U^{-1}W_{l}U\left\vert \vec{v}%
\right\rangle _{j_{l}}\left\vert 0\right\rangle .  \label{state 10}
\end{eqnarray}%
The word $W$ is $a_{X}^{\dagger }\{a_{Z}^{\dagger m_{1}}a_{Y}^{\dagger
m_{2}}\}$, $m_{\alpha }=0,1,...~$and so on. Here the curly bracket denotes
symmetrized states in the sector of $Z,Y,X$ with $U(3)$ global symmetry,
where the bracket denotes symmetrization over the global symmetry indices.
See e.g. \cite{Lewis-Brown:2020nmg,Pasukonis:2010rv}. The symmetrized states
itself is BPS. We make both the coherent state $F~$part and the word $W$
part BPS, hence the non-BPS energy is coming from the interaction between
the coherent state part and the word part. The above first example is a
special case of the second example for $m_{2}=0.$

In the following, we compute for the case (\ref{state 10}). The quartic
vertices that are relevant here are $\mathrm{Tr}|[Y,X]|^{2},\mathrm{Tr}%
|[Z,X]|^{2},\mathrm{Tr}|[Y,Z]|^{2}$. The anormalous dimension is due to the
interaction of impurity $a_{X}^{\dagger }$ with the $a_{Z}^{\dagger }$ and $%
a_{Y}^{\dagger }$ respectively, on the exponent. The interaction between $%
a_{Z}^{\dagger }$ and $a_{Y}^{\dagger }$ gives no anormalous dimension, due
to that they are in the exponent of the BPS coherent state, as calculated in
Sec 2. The word $W$ is by itself BPS because it is symmetrized for $%
a_{Z}^{\dagger }$, $a_{Y}^{\dagger }$, $a_{X}^{\dagger }$. Hence the
interaction between $Z,Y$ in this case does not lead to anormalous energy.
The $Z,Y$ composite here are BPS.

The Hamiltonian involved for the excitation energy is
\begin{equation}
\Delta =g^{2}~\mathrm{Tr}([a_{Y}^{\dagger },a_{X}^{\dagger
}][a_{X},a_{Y}])+g^{2}~\mathrm{Tr}([a_{Z}^{\dagger },a_{X}^{\dagger
}][a_{X},a_{Z}]).
\end{equation}%
Here, the interaction leading to the anormalous energy here are quartic
vertices between the $X$ in the word part and the $Z$ and $Y$, respectively
in the coherent state part. For example, $X$ in the word part is attached to
a quartic interaction vertex with a $Z$ in the coherent state $F$ part. And
after interaction, they Wick contract to the conjugate of the word part and
the conjugate of the coherent state part respectively. The coherent state
part itself is BPS as explained.

The method of calculation follows from \cite{Berenstein:2022srd}. One can
directly work with the quartic vertices as in BMN \cite{Berenstein:2002jq},
or use the dilatation operator \cite{Beisert:2003tq}. Here, $\left\vert
v_{y}\right\rangle _{\alpha }=\left\vert v_{z}\right\rangle _{\alpha }$ for $%
\alpha =i,j$,$~$as explained in Sec. 4.1, and we denote the pair of large
eigenvalues ${\vec{\lambda}_{i}}${, }${\vec{\lambda}_{j}}$. When acting on
the state, the extra annihilation operator $a_{Y}$ brings down $U\Lambda
_{Y}U^{-1}$, one to the left and the other to the right. These two pieces go
as
\begin{equation}
g^{2}\langle v_{y}|_{i}U^{-1}U\Lambda _{Y}U^{-1}[a_{X}^{\dagger
},a_{Y}^{\dagger }]U\left\vert v_{y}\right\rangle _{j}-g^{2}\langle
v_{y}|_{i}U^{-1}([a_{X}^{\dagger },a_{Y}^{\dagger }]U\Lambda
_{Y}U^{-1}U\left\vert v_{y}\right\rangle _{j}.
\end{equation}%
Since $\left\vert v_{y}\right\rangle _{i}$, $\left\vert v_{y}\right\rangle
_{j}~$are eigenstates of $\Lambda _{Y}$, we get
\begin{equation}
g^{2}(\lambda _{i}^{y}-\lambda _{j}^{y})\langle
v_{y}|_{i}U^{-1}[a_{X}^{\dagger },a_{Y}^{\dagger }]U\left\vert
v_{y}\right\rangle _{j}.  \label{a_Z_02}
\end{equation}%
We now proceed the process for computing with the conjugate vector, and $%
a_{Y}^{\dagger }$ brings down terms of $\bar{\Lambda}_{Y}$. We obtain an
integral that involves ${\tilde{U}}^{-1}$, so we get a factor $(\bar{\lambda}%
_{i}^{y}-\bar{\lambda}_{j}^{y})~$from (\ref{a_Z_02}). For $a_{Z}$, the
calculation is similar, and we get
\begin{equation}
g^{2}(\lambda _{i}^{z}-\lambda _{j}^{z})\langle
v_{z}|_{i}U^{-1}[a_{X}^{\dagger },a_{Z}^{\dagger }]U\left\vert
v_{z}\right\rangle _{j}.  \label{a_Y_02}
\end{equation}%
For the conjugate vector, $a_{Z}^{\dagger }$ brings down terms of $\bar{%
\Lambda}_{Z}$. Again, we have the integral and get a factor ($\bar{\lambda}%
_{i}^{z}-\bar{\lambda}_{j}^{z})$ from (\ref{a_Y_02}). Adding these two
pieces from (\ref{a_Z_02}) and (\ref{a_Y_02}), the spectrum is
\begin{equation}
H=\frac{g^{2}}{8\pi ^{2}}|\lambda _{z}^{i}-\lambda _{z}^{j}|^{2}+\frac{g^{2}%
}{8\pi ^{2}}|\lambda _{y}^{i}-\lambda _{y}^{j}|^{2}=\frac{g^{2}}{8\pi ^{2}}|{%
\vec{\lambda}_{i}-\vec{\lambda}_{j}|}^{2}.
\end{equation}%
The result subsumes the simplified case (\ref{state 08}).

The extra energy is perturbation energy around the eigenvalue background,
for a single pair of eigenvalues. We insert multiple words of this kind in (%
\ref{state 10}). Hence for inserting multiple words of this kind,%
\begin{equation}
H=\sum\limits_{(i,j)}\frac{g^{2}}{8\pi ^{2}}|{\vec{\lambda}_{i}-\vec{\lambda}%
_{j}|}^{2},  \label{spectrum_05}
\end{equation}%
where $(i,j)\in I_{s}$.$~$This is in agreement with alternative observations
\cite{deMelloKoch:2011ci,Carlson:2011hy,Berenstein:2013md}. The spectrum of
this type of Hamiltonian was also studied by \cite%
{Berenstein:2013md,Carlson:2011hy,deMelloKoch:2011ci,deMelloKoch:2012ck,Koch:2011hb}%
; see also related methods \cite{Cook:2007et}.

In this case, by the method of centrally extended algebra \cite%
{Beisert:2005tm},\cite{Berenstein:2014zxa}, the excitation energy are
written in a square-root form, whose expansion gives (\ref{spectrum_05}).
Hence the full square-root form is%
\begin{equation}
\sum\limits_{(i,j)}(\sqrt{1+\frac{g^{2}}{4\pi ^{2}}|{\vec{\lambda}}_{i}-{%
\vec{\lambda}}_{j}|^{2}}-1).  \label{spectrum_07}
\end{equation}%
Here we subtract the bare dimension out in (\ref{spectrum_07}) to obtain the
anormalous energy. The subtractions are the anormalous piece of the energy.
The centrally extended algebra gives important insights in \cite%
{Berenstein:2014zxa,deCarvalho:2020pdp,Gadde:2010ku} and these are in
agreement with the observations here.

These phenomena have very interesting dual interpretations on the gravity
side. This is string spectra in backreacted excited state spacetime
geometry. This analysis not only works for AdS background but also for
backreacted backgrounds, in the context of string excitations in backreacted
geometries. Eqn. (\ref{spectrum_07}) is in complete agreement with the
analysis and observations \cite%
{Berenstein:2020jen,Berenstein:2020grg,deMelloKoch:2016nxq,deMelloKoch:2018ert}
on both the gauge theory side and the gravity side. This spectrum is in
agreement with observations in the gravity side, in terms of magnon energy
in backreacted geometries, e.g. magnon energy in annularly shaped droplet
geometries.

The Eqn. (\ref{spectrum_05}) can be viewed as in the nonrelativistic regimes
of (\ref{spectrum_07}) and it was considered as the near BPS excitation of
heavy giant gravitons \cite{Carlson:2011hy}, in their near BPS limit. Eqn. (%
\ref{spectrum_07}) is the relativistic version of (\ref{spectrum_05}). This
near BPS limit would be the same as the one in spin-matrix theory \cite%
{Baiguera:2021hky,Harmark:2016cjq} and they are hence closely related.

\section{Discussion}

\label{sec_discussion}

The coherent state representation facilitates the calculation of ladder
operators and hence the dilatation operator. For instance, it converts
dilatation operator manipulations to algebraic manipulations. The coherent
state representation of the operators have the advantage of simplifying the
action of the above involved dilatation operators. We also checked higher
loop dilatation operators on quarter BPS coherent states and the action is
again vanishing as expected. By using nonrenormalization theorems, e.g. \cite%
{Lewis-Brown:2020nmg,Pasukonis:2010rv}, we infer that the BPS coherent
states are in the kernel of the anomalous dilatation operator. This
construction also works for coulomb branches.

In its original invention, \cite{Berenstein:2022srd} constructed coherent
states averaged over group orbit, for the purpose of gauge invariance. Here
we alternatively interpret this group average as a \textit{path integral} of
auxiliary variables coupled to the elementary letters of the theory. The
insertions of string words on the coherent state can be viewed as
operator-insertion in this path integral. We have generalized the word
insertion to multi-word insertions, by conveniently interpreting them as
operator insertions in this path integral. Hence, the group integral can be
viewed as a path integral of the auxiliary variables $U$ which are coupled
to the elementary letters and added string words.

We also constructed a meromorphic version of coherent states, which can be
transformed back to the full coherent states when $\det \Lambda _{Z}\neq 0$.
Here $G[T_{Y},\Lambda _{Z}]$ is a meromorphic version of $F[\Lambda
_{Y},\Lambda _{Z}]$. The meromorphic versions are particularly useful when
there are background $Z$ fields. We also constructed meromorphic version of
coherent states for product gauge groups and quivers. We give transformation
relations between different types of coherent states, among other things.

We constructed new type of coherent states in the $SL(2)$ sectors and
bosonic part of $PSU(1,1|2)$ sectors. These are operators with derivative
insertions. The large spin limit and small spin limit can be reached by
varying the coherent state amplitudes. We have also discussed other new
states related to them.

The different coherent states differ by the symmetry properties of the
multi-parameters of the coherent states. The coherent states constructed in
\cite{Berenstein:2022srd} have manifest permutation symmetries among the
eigenvalues. Other coherent states have gauge-fixed these symmetries \cite%
{Lin:2017dnz}, and can be viewed as gauge-fixed versions. Although different
construction of coherent states may have different norms, there are many
norm-independent features and observables of coherent state construction,
for example, when calculating the vev of $a_{Z}^{\dagger n}a_{Z}^{l}$, the
norm factor of the numerator and of the denominator cancels each other and
is hence norm-independent.

We have used these coherent states to describe BPS and near BPS states, in
gauge-string correspondence. We analyzed string-added coherent states, in
general a product of a coherent state and a word, and calculated anormalous
dimension energy. The string-added coherent state captures a class of near
BPS spectra. One can use more general words, by using BMN like operators or
spin chain like operators. The method of \cite{Berenstein:2022srd} gives
great insights for computing excitations on supersymmetric backgrounds.
Magnon excitations have also been computed for both gauge theory side and
gravity side in related works.\ The agreement to these calculations is
astounding, because the underlying methods and physical perspectives of these
different approaches are different and alternative to each other. This
approach is also intimately related to the effective eigenvalue model. In
these involved circumstances, the eigenvalues provide some string theory
background, and the off-diagonal fluctuations provide string excitations
\cite{Berenstein:2005aa}. Non-BPS excitations are added as a perturbation of
the BPS excitations. These observations pointed out to a way to solve the
strong coupling dynamics of these near-BPS systems, and we have encountered
some new strong coupling phenomena in those regimes. On a very different
perspective, the approach of adding excitations may be relevant for \cite%
{Morozov:1992zb,Kazakov:1992ym}.

On the gravity side, these coherent states are related to giant gravitons
and backreacted geometries. The individual eigenvalue of coherent state is
the collective coordinate of giant gravitons \cite{Berenstein:2022srd}, see
also related discussions \cite{Berenstein:2017abm,Lin:2017dnz}. The coherent
state construction is an important step in this regime.\ These are regimes
where the single-trace and multi-trace operators are very difficult to
handle efficiently. The effective action of the coherent state collective
coordinate in the Lagrangian formalism has been constructed \cite%
{Berenstein:2022srd}, where properties of giant gravitons and geometric
quotient and projection of the transverse space in the dual gravity
geometry, in the case of quiver theory, have been observed by this approach
\cite{Berenstein:2022srd}. These are important for giant gravitons and
backreacted geometries and related to various important discussions of e.g.
\cite{Biswas:2006tj}-\cite{Pasukonis:2012zj}.

Our approach shield new lights to near BPS states and near BPS sectors. We
conveniently look at eighth BPS states and near BPS states, in this set-up.
Another way to understand near BPS states is by spin-matrix theory e.g. \cite%
{Baiguera:2021hky,Harmark:2016cjq}, which is closely related to the
approaches in this paper. The spin-matrix theory is very insightful for both
near BPS string states and near BPS giant gravitons. This is intimately
related to the consistent reductions. The related consistent reductions are
also very interestingly discussed in \cite{Ishiki:2006rt} to fewer
dimensions.

The involved path integral can be computed by saddle point method and it is
an exact computation in the context of localization. The localization
technique is also very useful for Wilson operators. Localization methods
have occurred remarkably in Wilson loops in e.g. \cite{Pestun:2009nn} and
related references. Moreover, emergent geometries are also dual to large
Wilson loop operators, e.g. \cite{Yamaguchi:2006te}-\cite{Gomis:2008qa} as
some examples. It would be good to have an unified understanding together
with these circumstances.

The coherent state operators, Young tableau operators, and large Wilson
operators are heavy operators. They have gravitational dual descriptions.
The gravitational dual system to these very heavy operators, involves
backreacted emergent geometries, e.g. \cite{Gava:2006pu}-\cite{Mukhi:2005cv}
as some examples, and see their related references. These heavy states
induce backreactions in the dual quantum gravity system. The coherent states
are also heavy excited states. The coherent state approach leads to higher
multi-pole moments in the gravity side, which can be measured on the
boundary of the gravity side, as discussed in Sec. 2. Our scenarios are
closely related to fuzzball geometries and their related discussions, e.g.
\cite{Mathur:2005ai}-\cite{Chakrabarty:2021sff} and \cite{Belin:2020zjb}.

Near BPS states can also describe near BPS black holes, which are related to
various giant configurations and intersecting giants. The string
configurations \cite{Berenstein:2022cju} on giants are relevant. A deeper
understanding of the quarter and eighth BPS sector as well as near BPS
sector will lead to implications for physics of extremal and near-extremal
black holes e.g. \cite{Balasubramanian:2007bs,Fareghbal:2008ar}.

The $SO(N)$ and $Sp(N)$ cases of this approach have been worked out in \cite%
{Holguin:2022drf}. The related Young tableaux basis have been addressed and
developed in details in e.g. \cite%
{Caputa:2013hr,Caputa:2013vla,Lewis-Brown:2018dje} and their related
references. The case of other gauge groups and product gauge groups have
also been considered \cite{Holguin:2022drf,Berenstein:2022srd}, as well as
in this paper. This approach also works for $SO/Sp$ theories, quiver
theories, and coulomb branches. Although the analysis of the current paper
is mainly for $U(N)$ and $SU(N)$ case, many features can be similar in the
case of these other gauge groups.

\section*{Acknowledgments}

We would like to thank B. Czech, R. de Mello Koch, L.-Y. Hung, Y. Jiang, S.
Lin, Y. Liu, S. Ramgoolam, J. Simon, M. Sperling, R. Suzuki for useful
discussions or communications. The work was supported in part by National
Key R\&D Program of China grant No. 2020YFA0713000, by Overseas high-level
talents program, and by Fundamental Research Funds for the Central
Universities of China.

\end{document}